\documentclass[letter,11pt]{article}
\usepackage{amsfonts}

\newtheorem{lemma}{Lemma}
\newtheorem{theorem}{Theorem}
\newenvironment{proof}%
{\begin{trivlist}\item[\hspace*{\labelsep}{\it Proof.\/}]}%
{\hfill$\Box$\end{trivlist}}
\newenvironment{proof1}%
{\begin{trivlist}\item[\hspace*{\labelsep}{\it \/}]}%
{\hfill$\Box$\end{trivlist}}

\newcommand{\para}{\medskip\noindent}

\newcommand{\head}[1]
 {\markright{\hbox to 0pt{\vtop to 0pt{\hbox{}\vskip 3mm \hrule
 width  \textwidth \vss} \hss}{\sc #1}}}
\usepackage{latexsym,graphicx}
\begin{document}
\title{\bf  New Upper Bounds on The Approximability of 3D Strip Packing } 

\author{Xin Han$^1$ \hspace{3mm} Kazuo Iwama$^1$\hspace{3mm}
Guochuan Zhang$^{2}$
\\ {\small $^1$ School of Informatics, Kyoto University, Kyoto
606-8501, Japan} \\ {\small \{hanxin,
iwama\}@kuis.kyoto-u.ac.jp}
\\ {\small $^2$ Department of Mathematics, Zhejiang University, China}
\\ {\small zgc@zju.edu.cn}}
\date{}
\maketitle

\begin{abstract}
In this paper, we study the 3D strip packing problem in which we are
given a list of 3-dimensional boxes and required to pack all of them
into a 3-dimensional strip with length 1 and width 1  and unlimited
height to minimize the height used. Our results are below: i) we
give an approximation algorithm with asymptotic worst-case ratio
$1.69103$, which improves the previous best bound of $2+\epsilon$ by
Jansen and Solis-Oba of SODA 2006; ii) we also present an asymptotic
PTAS for the case in which all items have {\em square} bases.
\end{abstract}

\section{Introduction}
\begin{figure}[htbp]
 \begin{center}
  \includegraphics[scale=0.6]{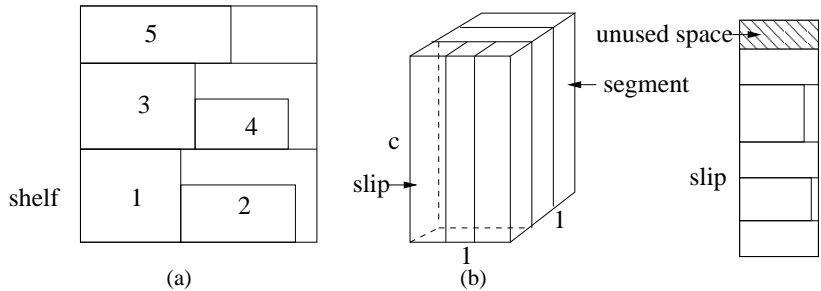}
  \caption{ Shelves and segments}
  \label{fig:ss}
  \end{center}
\end{figure}

For packing 2D items into bins  or a strip, it is a natural idea to
exploit techniques for packing lower dimensional (i.e., 1D) items.
The {\em two-stage packing} is particularly well-known: As shown in
Fig.~\ref{fig:ss} (a), a bin (or a strip) is divided into {\em
shelves} and each shelf contains a single layer of items. After
packing items into shelves, the problem of packing shelves into bins
(or a strip) obviously becomes the 1D bin (or strip) packing
problem. The idea originally comes from cutting a large board into
smaller items efficiently \cite{GG65}; one can cut the board only in
two stages, i.e., cutting horizontally first and  then vertically.

It should be noted that many existing 2D packing algorithms
\cite{CGJ82,CGJT80,Cap02} are based on this two-stage packing. In
2002, Caprara \cite{Cap02} established the relation between 2D Bin
Packing (2BP) and 2D Shelf Bin Packing (2SBP). Namely the maximum
ratio between the optimal cost for 2SBP and that for 2BP is equal to
$T_{\infty} = 1.691...$ which is the well-known approximation factor
of the Harmonic algorithm for 1D Bin Packing \cite{LL85}. (A similar
relation between 2D Strip Packing (2SP) and 2D Shelf Strip Packing
(2SSP) was also established by Csirik and Woeginger \cite{CW97}.) As
an important byproduct, Caprara also showed an approximation
algorithm for 2BP whose asymptotic worst-case ratio is arbitrarily
close to $T_{\infty}$, which first broke the barrier of two for the
upper bound on the approximability of this problem.

\para{\bf Our contribution} This paper extends the two-stage packing
into the 3D Strip Packing (3SP) and obtains an approximation
algorithm whose asymptotic worst-case ratio is arbitrarily close to
$T_{\infty}$. Our model is standard (see
Section~\ref{sec:definition} for details) and the previous best
bound is $2+\epsilon$ by Jansen and Solis-Oba \cite{JS06}. We also
show that there is an APTAS for the special case in which all items
have square bases.

Our algorithms use a {\em segment} as shown in Fig.~\ref{fig:ss} (b)
instead of a shelf in the 2D case. For packing items (whose three
sides are all at most 1.0) into a segment, we first divide a segment
into slips and pack the items into slips by  the next-fit
(NF)algorithm. The key idea is to make the height $c$ of each
segment sufficiently large (within a constant), which effectively
kills the inefficiency of the algorithm for the vertical direction
in the sense that the unused space at the top of the segment is
relatively small. After packing items into segments of the fixed
height (=c) and fixed length (=1.0), we can obviously use a
one-dimensional bin packing algorithm to pack segments.

\para{\bf Previous results:}
On 3D Strip Packing, Li and Cheng \cite{LC90} presented the first
approximation algorithm with asymptotic worst case ratio 3.25. Two
years later, they gave an online algorithm for the problem with
asymptotic worst-case (competitive) ratio arbitrarily close to
$(1.69103...)^2 \approx 2.8596$ \cite{LC92}. Then Miyazawa and
Wakabayashi  \cite{MW97,MW04} improved the asymptotic worst-case
ratio to 2.67 and 2.64. Very recently, Jansen and Solis-Oba
\cite{JS06} improved the asymptotic worst-case ratio to
$2+\epsilon$.

On 2D Strip Packing, Coffman et al.~\cite{CGJT80} presented
algorithms based on NFDH (Next Fit Decreasing Height) and FFDH
(First Fit Decreasing Height), and showed that the respective
asymptotic worst-case ratios are 2 and 1.7. Golan \cite{Golan81} and
Baker et al. \cite{BBK81} improved the bound to $4/3$ and $5/4$,
respectively. Using linear programming and randomization techniques,
an asymptotic fully polynomial time approximation schemes (AFPTAS)
was given by Kenyon and R\'emila \cite{KR00}.

On 2D Bin Packing, in 1982, Chung, Garey and Johnson \cite{CGJ82}
presented an approximation algorithm with asymptotic worst-case
ratio at most 2.125. Caprara \cite{Cap02} improved the upper bound
to $1.6910...$. On the other hand, Bansal et al. \cite{BS06} showed
that the 2D bin packing problem does not admit an APTAS. Chleb\'ik
and Chleb\'ikov\'a~\cite{CC06} further gave an explicit lower bound
$1+\frac1{2196}$. Since the 2D bin packing problem is a special case
of the 3D strip packing problem, the lower bound holds for 3D strip
packing too.

\section{Problems and Notations}
\label{sec:definition} Our model is exactly the same as \cite{JS06}.
Given an input list $L$ of $n$ three-dimensional boxes, in which
each box has  length, width and height at most 1 respectively, 3SP
is to pack all boxes into a  3D strip  (rectangular parallelepiped)
of width 1, length 1 and minimum height, so that the boxes do not
overlap. In this paper we consider the orthogonal version of the
problem without rotations, i.e., the boxes must be packed  so that
their faces are parallel  to the faces of the strip and the boxes
are oriented and cannot be rotated. The problem is obviously
NP-hard. For approximation algorithms, we use the standard measure
to evaluate them, i.e., the worst-case ratio. In this paper, we
consider the asymptotic worst-case ratio. Given an input list {\em
L} and an approximation algorithm $A$, we denote by $OPT(L)$ and
$A(L)$, respectively, the height
 used by an optimal  algorithm and the height
by  algorithm $A$ for list $L$.
The {\em asymptotic worst-case ratio} $R_A^{\infty}$
of algorithm $A$ is defined by
\[
     R_A^{\infty} =\lim_{n \to \infty} \sup_{L}\{ A(L)/OPT(L)| OPT(L) = n\}.
\]

\section{Basic tools for algorithms and their analysis}
\label{sec:binpacking}

\para {\bf Fractional Bin Packing} (FBP). The continuous version of
bin packing plays an important role in designing an asymptotic PTAS
\cite{VL81,KK82}. We first give its definition and some properties.
Given an instance $I$ of one dimensional bin packing, suppose that
there are $p$ {\em distinct} sizes of the items in $I$, where $p$ is
a constant. Let $s_1 > s_2 > ... > s_{p}$ be the distinct item sizes
and $n_j$ be the number of items of size $s_j$ for $j=1,\ldots, p$.
A {\em feasible pattern} is a vector $v=(v_1,\dots,v_p)$  such that
$\sum_{j=1}^{p} v_{j}s_j \le 1$, i.e., all items in a  feasible
pattern would fit in one bin. Let $\nu$  denote the collection of
all feasible patterns for $I$ and $v^{i} = (v_1^i,\ldots, v_p^i)$
denote the $i$-th pattern in $\nu$, where $v_j^i$ is the number of
items of size $s_j$ in the $i$-th pattern. We further denote $x_i$
to be the number of bins being needed for packing the $i$-th
feasible pattern in $\nu$. If we allow $x_i$ to be a fractional
number, then the problem becomes the fractional bin packing problem
(FBP) and corresponds to the following {\em Linear Program} (LP):

\begin{eqnarray}
       \textrm{Min } &  \sum_{v^i \in \nu} x_i \nonumber &  \\
       \textrm{s.t. }   &  \sum_{v^i \in \nu} v_j^{i}  x_i \ge n_j, &
                             j = 1,\dots,p \label{eqn:primary} \\
                               &         x_i \ge 0,  &  v^{i} \in \nu. \nonumber
\end{eqnarray}

\noindent  The {\em LP dual} of (\ref{eqn:primary}) is given as follows:
    \begin{eqnarray}
       \textrm{max.  }  & \sum_{j =1}^{p} n_j  \pi_{j}  \nonumber &\\
       \textrm{s.t.  }  & \sum_{j=1}^{p} v_j^{i} \pi_j \le 1, \textrm{  }
                          &   v^i \in \nu  \label{eqn:dual}\\
                                      &  \pi_j \ge 0, &  j = 1,\dots,p . \nonumber
  \end{eqnarray}

Optimal values for (\ref{eqn:primary}) and (\ref{eqn:dual}) coincide
and the following important lemma is due to \cite{Cap02},
 \begin{lemma}
  There exists an optimal solution $\pi^{*}$ of (\ref{eqn:dual}) such that
  $\pi_{1}^{*} \ge \pi_{2}^{*} \ge  \dots \ge \pi_{p}^{*}$
 (recalling $s_1 > s_2 > \dots > s_{p}$).
 \label{lemma:order}
 \end{lemma}

The following lemma \cite{Cap02, VL81}, says that the optimal values
for BP and FBP are almost equal.
\begin{lemma}\label{lemma:Fractional}
For any bin packing instance $I$ and for any $\epsilon > 0$, we have
$ OPT_{BP}(I) \le (1 + \epsilon) OPT_{FBP}(I) + O(\epsilon^{-2})$,
where $OPT_{FBP}(I)$ is the optimal value for FBP.
\end{lemma}

\para{\bf Harmonic algorithm.}
The Harmonic algorithm was introduced by Lee and Lee~\cite{LL85}.
Given a (one-dimensional) bin packing instance $I$ and an integer $
k>0$, we say an item $i$ belongs to type $t$ if its size $s_i \in
(\frac{1}{t+1}, \frac1t]$ for $t = 1,\dots,k-1$ and to type $k$ if
$s_i \in (0, \frac1k]$, where $k$ is a constant. Then the Harmonic
algorithm packs items of different types into different bins. During
packing, if the current item of type $t$ does not fit in the
corresponding bin, then the algorithm closes the bin and opens a new
one. Given an item of size $x$, we define a weighting function
$f_k(x)$ as follows:
\begin{displaymath}
   f_k(x) = \left\{ \begin{array}{ll}
                     \frac1t, & \textrm{ if $\frac1{t+1} < x \le \frac1t$ with
       $1 \le t < k$,}  \\
                    \frac{kx}{k-1} & \textrm{ if $0 < x \le \frac1k$.}
                    \end{array}
            \right.
\end{displaymath}
Let $t_1 =1$, $t_{i+1}=t_i(t_i +1)$ for $i\ge 1$. For a positive
integer $k$, let $m(k)$ be the integer such that $t_{m(k)} < k \le
t_{m(k)+1}$. $T_k = \sum_{i=1}^{m(k)} \frac1{t_i} +
\frac1{t_{m(k)+1}} \cdot \frac{k}{k-1}$. Note that $T_{\infty}
=\lim_{k \to \infty} T_k \approx 1.69103 $

The weighting function $f_k(x)$ satisfies the following property
(see \cite{LL85}):
  \begin{lemma}
   For each sequence $x_1,\dots,x_m$ with $x_i \in (0,1]$ and
  $\sum_{i=1}^{m} x_i \le 1$,
   \[
       \sum_{i=1}^{m} f_k(x_i) \le T_k.
   \]
   \label{lemma:Harmonic}
  \end{lemma}
\para {\bf NFDH packing.}
{\em NFDH} was first proposed by Meir and Moser \cite{MM68} for
packing a set of {\em squares} into a rectangular bin, but {\em
NFDH} packing can also be applied to pack rectangles. It simply
works as follows. First sort all rectangles in non-increasing order
of their heights. Then pack them into the bin level by level and in
each level we use the Next Fit (NF) algorithm. If a level cannot
accommodate the current rectangle, then we close it  (will never be
used again) and open a new one. (see Figure ~\ref{fig:NFDS} (c)).
Note that NFDH packing can be extended for multidimensional
packing~\cite{MM68, BS06}.

\section{ New upper bounds for 3D strip packing  }
We call our algorithm 3D Segment Strip Packing (3SSP).


\subsection{ Algorithm 3SSP }
Given an item $R=(l, w, h)$, where $l$, $w$ and $h$ are its length,
width and height respectively, we may use $l(R)$, $w(R)$ and $h(R)$
to denote the three parameters as well. Algorithm 3SSP has the
following 3 main steps.

\begin{enumerate}

\item Divide all items into $k$ groups $G_1,G_2,\dots,G_k$
        such that those in $G_i$
        have their lengths  in range $(\frac1{i+1},\frac1i]$,
        where $k$ is a constant.
\item Sort all $G_i$-items by their width such that
        $G_i=(R_1, R_2, \ldots, R_{n_i})$ and
        $w(R_1) \ge w(R_2) \ge \cdots \ge w(R_{n_i})$,
        where $n_i$ is the number of items in group $G_i$
        for $1 \le i \le k$. Then pack all items in $G_i$, for $1\le
        i<k$, into segments by algorithm GNF (given later).
        For $i =k$, pack all items in $G_k$
        into segments by algorithm GNFDH (given later).
\item  When all items in group $G_i$ have
         been packed into segments,
         for $1 \le i \le k$,
         then regard all segments as one dimensional items
         (Only their width is considered) and call an asymptotic PTAS for
         one-dimensional bin packing (e.g. \cite{KK82, VL81}) to pack these segments.
\end{enumerate}

In the following we give the procedures to pack 3D items into
segments, which are the cores of algorithm 3SSP. We deal with $G_i$
items ($1\le i<k$) and $G_k$ items separately.

\para{\bf GNF:} Consider $G_i$ items ($1\le i<k$). Given
$G_i=(R_1, R_2, \dots, R_{n_i})$ such that $w(R_1) \ge w(R_2) \ge
\cdots \ge w(R_{n_{i}})$.

\begin{figure}[htbp]
\begin{center}
\includegraphics[scale=0.5]{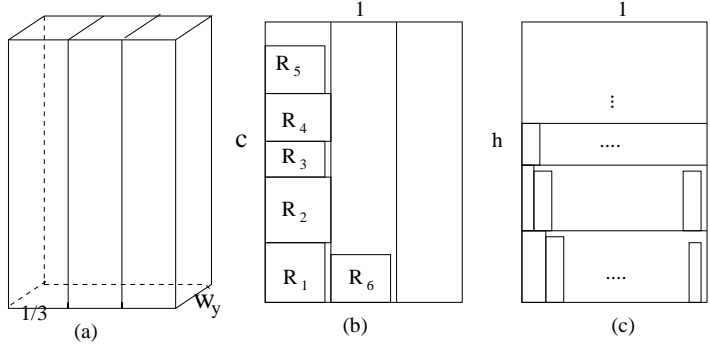}
\caption{GNF for $G_3$; the projection of GNFDH packing without
considering widths} \label{fig:NFDS}
\end{center}
\end{figure}

\begin{enumerate}
\item Open a new segment of size $(1,w_y,c)$, where  $w_y \gets w(R_1)$.

\item    Divide this segment into $i$ pieces of
         slips of sizes $(\frac1{i},w_y,c)$, as shown in
          Figure ~\ref{fig:NFDS} (a),
         then without considering their widths and lengths,
         pack items into these slips by Next Fit.
         (see Figure ~\ref{fig:NFDS} (b)).

\item  If there are remaining items, re-index them and
         go to Step 1.
\end{enumerate}

\para {\bf GNFDH:}
Given $G_k=(R_1, R_2, \dots, R_{n_k})$ such that $w(R_1) \ge w(R_2)
\ge \cdots \ge w(R_{n_{k}})$.

\begin{enumerate}
\item  Open a new segment with size $(1,w_y,c)$, where $w_y  \gets w(R_1)$.

\item  Find a maximal index $j$ such that $R_1,R_2,\dots,R_j$
         can be placed into the segment by NFDH without considering
         their widths. Pack the $j$ items by NFDH.
         (See Figure ~\ref{fig:NFDS} (c))

\item  Re-index the remaining items in $G_k$ (if any), go to Step
  1.

\end{enumerate}

\subsection{ Analysis of the algorithm}\label{subsec:Dual}

In the algorithm analysis, {\em dual feasible functions} by Fekete
and Scheper \cite{FS01} play a crucial role. (Similar notions are
used as weighting functions \cite{GGJ76,GGU71,LL85,S02,SS03})
Suppose that a function  $f : [0,1] \to [0,1]$ satisfies
$\sum_{i=1}^{m} f(x_{i}) \le 1$  for any sequence  $x_1,\ldots,x_m$
such that $\sum_{i=1}^{m} x_{i} \le 1$ and $x_{i} \in [0,1]$. Then
$f$ is called a {\em dual feasible} function. Here are two specific
examples: Let  $\bar{\pi} = (\bar{\pi_1},...,\bar{\pi_p})$ be a
feasible solution of (\ref{eqn:dual}) (dual LP for FBP in Section
\ref{sec:binpacking})
 satisfying the requirement of Lemma \ref{lemma:order} and
 let $\bar{\pi}_{p+1}:=0$, $s_0 :=1$ and $s_{p+1} := 0$.
 Define a new function $g$ by
   \[
 g(0)=0, \textrm{ and } g(x)=  \bar{\pi}_{j}, \textrm{ for } x \in [s_j, s_{j-1}).
  \]

The other example is $f_k$ defined in Section~\ref{sec:binpacking}.

\begin{lemma}\label{lemma:feasible} \cite{Cap02}
Both $g(x)$ and $\frac{f_k(x)}{T_k}$ are dual feasible functions.
\end{lemma}

Using these two functions, we define the {\em modified volume}
$W(R)$ of an item  $R=(l, w, h)$ as
\[
W(R)=f_k(l)\cdot g(w) \cdot h.
\]
The total modified volume of the input list $L$ of items is $W(L) =
\sum_{R \in L} W(R)$.

We need one more lemma regarding dual feasible functions and 2D
packing: let $(l_1,w_1),\ldots,(l_m,w_m)$ be 2D items which can be
packed into a square bin of size (1,1), and $f_1$ and $f_2$ be dual
feasible functions. Then we have the following lemma (see
\cite{CV93,SS03} for the proof), which is important for bounding the
total modified volume.

\begin{lemma}\label{lemma:TwobBound}
$\sum_{i=1}^{m} f_1(l_i) f_2(w_i) \le 1 $.
\end{lemma}
Now, we are ready to prove the upper bound for the approximability
of our algorithm 3SSP. Let $I(L)$ be the 1-dimensional item list
obtained after Step 2 of 3SSP, i.e., the list of the widths of the
segments. Recall that $c$ is the height of the segment and $k$ is
the parameter of the Harmonic algorithm. Let $OPT_{BP}(I(L))$ be the
optimal cost of 1-dimensional bin packing for the list $I(L)$ and
$OPT(L)$ be the optimal cost for 3D Strip Packing for the list $L$.
Our goal is thus to prove the following theorem.

\begin{theorem} \label{theorem:Tk}
For any $\epsilon > 0$, $c \cdot OPT_{BP}(I(L)) \le \frac{c}{c-1}
(1+\epsilon) T_k  OPT(L) + O(ck\epsilon^{-2})$.
\end{theorem}
Since we employ some APTAS for packing $I(L)$, algorithm 3SSP
achieves the cost arbitrarily close to $c\cdot OPT_{BP}(I(L))$ in
the asymptotic case. It shows that the asymptotic worst-case ratio
of 3SSP is at most $\frac{c}{c-1} (1+\epsilon) T_k $ for any given
$\epsilon>0$, which tends to $T_{\infty}$ as $\epsilon \to 0$ and
the constants $c$ and $k$ take sufficiently large integers.

The basic idea of the proof is to establish the relation of the left
and right-hand sides of the inequalities in the theorem to the total
modified volume. Recall that 3SSP uses different segments for each
$G_i$. A segment is called type $i$ if it contains $G_i$ items. For
$q= 1,\dots,k$, let $m^{q}$ be the number of segments of type $q$
and $w_i^{q}$ the width of the $i$-th segment of type $q$ , where $1
\le i \le m^q$. By algorithm 3SSP, we have

\begin{equation} \label{eqn:widths}
w_1^{q} \ge w_2^{q} \ge \cdots \ge w_{m^{q}}^{q}.
\end{equation}
Noting that $g(\cdot)$ is the function defined in Subsection
\ref{subsec:Dual} for instance $I(L)$, by Lemma \ref{lemma:order},
we have
\begin{equation}
  g(w_1^{q}) \ge  g(w_2^{q}) \ge \cdots \ge g(w_{m^{q}}^{q}).
  \label{eqn:weights}
  \end{equation}
Let  $G^{q} := \sum_{i=1}^{m^{q}}g(w_{i}^{q})$ denote the total
modified width of the segments of type $q$. Now, we give a lower
bound for $W(L)$. For convenience, we define $w_{m^{q}+1}^{q} = 0$
for all $q$'s.

\begin{lemma}  \label{lemma:allslip}
The total modified volume $W(L) > (c-1) \sum_{q=1}^{k} G^q - ck$.
\end{lemma}
\begin{proof}
Let $S_{i}^{q}$ be the $i$-th segment of type $q$ and $L_{i}^{q}$ be
the set of all boxes in $S_{i}^{q}$. We first prove that
\begin{equation}\label{eqn:subslip}
 W(L_{i}^{q}) \ge  (c-1)g(w_{i+1}^{q}).
\end{equation}

\vskip 2mm \noindent Case 1. $q \ne k$. If $i=m^{q}$, $W(L_{i}^{q})
\ge (c-1)g(w_{i+1}^{q})$ since $g(w_{m^{q}+1}^q) = 0$ and
$W(L_{m^{q}}^{q}) \ge 0$. Otherwise, i.e., $ 1 \le i  < m^{q}$, by
GNF packing, we have the height packed in every slip of $S_{i}^{q}$
is at least $(c-1)$ and every box in $S_{i}^{q}$ has length in
$(\frac1{q+1}, \frac1q]$ and width at least $w_{i+1}^{q}$ (i.e., the
width of the next segment). Remember $f_k(x) = \frac1q$ where $x$ in
$(\frac1{q+1}, \frac1q]$. By (\ref{eqn:weights}), the total weight
of the boxes in every slip is at least
 \[
  \frac1q \cdot g(w_{i+1}^{q}) \cdot (c-1) =  \frac{c-1}{q}\cdot g(w_{i+1}^{q}).
\]
  Since  there are $q$ slips in segment $S_{i}^{q}$,
  \[W(L_{i}^{q}) \ge q \cdot \frac{c-1}{q} \cdot g(w_{i+1}^{q}) = (c-1) g(w_{i+1}^{q}).\]

\vskip 2mm \noindent Case 2. $q = k$. If $i = m^{q}$, $W(L_{i}^{q})
\ge (c-1)g(w_{i+1}^{q})$ still holds. Otherwise, i.e., $ 1 \le i <
m^{q}$. Consider GNFDH packing for items of type $k$. Assume there
are $l$ levels in $S_{i}^{k}$, and their heights are
$h_1,h_2,\dots,h_l$, respectively. Set $h_{l+1} = c - \sum_{j=1}^l
h_i$. By NFDH packing, if $i < m^q$, we have $h_1 \ge h_2 \ge \dots
\ge h_{l+1}$ and every box in  $S_{i}^{k}$ has width at least
$w_{i+1}^{k}$,  length at most $\frac1k$. Hence the total sum of
lengths in every level is at least $1-\frac1k$. Remember $f_k(x) =
\frac{kx}{k-1}$ where $x$ in  $[0, \frac1k]$. And every box in the
$j$-th level of $S_{i}^{k}$ has height at least $h_{j+1}$. So the
total weight in the $j$-th level is at least
 \[
     h_{j+1} \cdot g(w_{i+1}^{k})\sum_{R} \frac{k}{k-1}l(R)
     \ge h_{j+1} \cdot g(w_{i+1}^{k}).
\]
 Since $\sum_{j=1}^l h_j \ge c-1$, the total weight of $L_i^{k}$ is
  \[
     W(L_i^{k})  \ge \sum_{j=1}^l g(w_{i+1}^{k}) h_j  \ge (c-1)g(w_{i+1}^{k}).
 \]
So the inequality (\ref{eqn:subslip}) holds.

Since there are $k$ types of segments and in every type $q$ there
are $m^{q}$ segments, so
\begin{displaymath}
       W(L) = \sum_{q=1}^{k} \sum_{i=1}^{m^{q}} W(L_i^{q}).
     \end{displaymath}
 By (\ref{eqn:subslip}),
   \begin{displaymath}
   \begin{array}{ccc}
   W(L) &\ge& (c-1)\sum_{q=1}^{k} \sum_{i=1}^{m^{q}} g(w_{i+1}^q) \nonumber \\
        & = &    (c-1)\sum_{q=1}^{k} (G^q - g(w_{1}^q)) \nonumber \\
        & > &  (c-1)\sum_{q=1}^{k} G^q - ck.
   \end{array}
   \end{displaymath}
The last inequality follows directly from $ g(w_{1}^q) \le 1$.
\end{proof}

Next we give an upper bound for total modified volume $W(L)$.
\begin{lemma}\label{lemma:integrate}
Given any input list $L$ over $[0,1]^{3}$, the total modified volume
$W(L) \le T_k OPT(L)$.
\end{lemma}
\begin{proof}
Consider an optimal packing for an input list $L$. For each item of
$L$ we draw two horizontal planes at its bottom and top, shown as
Figure ~\ref{fig:layers}.
  \begin{figure}[htbp]
  \begin{center}
  \includegraphics[scale=0.5]{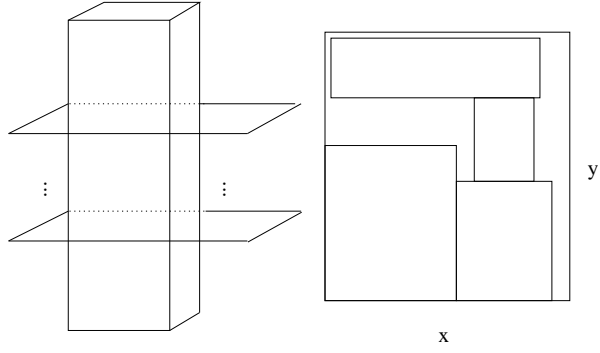}
  \caption{Cutting an optimal packing  and a layer projection on xy-plane}
  \label{fig:layers}
  \end{center}
\end{figure}
These planes cut the optimal packing into layers such that all items
(may be part of the original items) in a layer have the same height.
Then we can see that each layer is associated with a {\em feasible
packing} on a square bin of $(1,1)$ by ignoring the heights. Assume
that after cutting, totally, there are $l$ layers and their heights
are $\delta_{1},\delta_{2},\dots,\delta_{l}$, respectively. By
Lemmas \ref{lemma:Harmonic} and \ref{lemma:feasible}, we have
$\sum_{x \in S} f_k(x) \le T_k$ and $\sum_{x \in S} g(x) \le 1$ for
any list $S$ with  $\sum_{x \in S} x \le 1$. Since in the $i$-th
layer, every item has height $\delta_{i}$, by Lemma
\ref{lemma:TwobBound}, the total weight of all items in  the $i$-th
layer is at most
  \[
  T_k \times 1 \times \delta_{i}.
   \]
   Since
  \[
     \sum_{i=1}^{l}\delta_{i} = OPT(L),
  \]
   then
   \[
    W(L) \le \sum_{i=1}^{l} \delta_{i}T_k = T_k \cdot OPT(L).
  \]
 \end{proof}

Now it is straightforward to prove Theorem \ref{theorem:Tk}.
\begin{proof1} (Proof of Theorem \ref{theorem:Tk})
By Lemma \ref{lemma:Fractional}, we have
  \[
  OPT_{BP}(I(L))  \le (1 + \epsilon)OPT_{FBP}(I(L)) + O(\epsilon^{-2}).
  \]
By the duality of FBP and the dual FBP, as used in \cite{Cap02}, we
have
 \[
 OPT_{FBP}(I(L)) = \sum_{q=1}^{k} G^{q},
  \]
where  $G^{q} := \sum_{i=1}^{m^{q}}g(w_{i}^{q})$ denotes the overall
modified width of the segments of type $q$. By Lemmas
\ref{lemma:allslip} and \ref{lemma:integrate},
\begin{eqnarray*}
  c \cdot OPT_{BP}(I(L)) & \le & c (1 + \epsilon)OPT_{FBP}(I(L)) + O(c\epsilon^{-2}) \\
                   &= & c(1+\epsilon)\sum_{q=1}^{k} G^{q} + O(c\epsilon^{-2}) \\
                   &< & c(1+\epsilon) \frac{W(L)+ck}{c-1} + O(c\epsilon^{-2}) \\
                   &\le& \frac{c}{c-1}(1+\epsilon)T_k  OPT(L) +  O(ck\epsilon^{-2}).
\end{eqnarray*}
\end{proof1}

\noindent{\bf Remark.} Our algorithm can also be applied to the
parametric case in which the boxes have bounded length (or width).
Then by Theorem \ref{theorem:Tk} the asymptotic worst-case ratio in
the parametric case that all boxes have width or length bounded from
above by $\alpha$ is stated in the following table, which is better
than the previous parametric ratio $R_{para}^{\infty}$ in
\cite{MW05}.

\begin{center}
\begin{tabular}{|c|c|c|c|c|}
  \hline
   $ \alpha \in $ & ($\frac12$,1] & ($\frac13$, $\frac12$] & ($\frac14$, $\frac13$] & ( $\frac15$, $\frac14$] \\
  \hline
  $ R_{3SSP}^{\infty}$ & 1.691... & 1.423... & 1.302... & 1.234...  \\
  \hline
   $R_{para}^{\infty}$         & 3.050... & 2.028... & 1.684... & 1.511...  \\
  \hline
 \end{tabular}
 \end{center}

\section{APTAS for packing items with square bases}
In this section, by combining the techniques for 2D strip packing
\cite{KR00} and 2D bin packing \cite{BS06}, we give an APTAS for the
case that the boxes have square bases (bottoms).

The standard ideas in our scheme are below:
\begin{itemize}
\item Create a gap between large items and small items such that the
items fall into the gap do not affect the packing significantly.

\item Pack large items in the way similar to 2D strip packing \cite{KR00} and
pack the other items by NFDH \cite{MM68,BS06,JS06}.

\end{itemize}
We use a multidimensional version of NFDH in \cite{MM68,BS06},
called MNFDH, to pack items with small base sizes into a 3D bin or a
strip. The lemma below can be obtained directly from \cite{MM68}
(see also \cite{BS06, JS06}).

\begin{lemma}\label{lemma:MNFDH}
Let $I$ be a set of 3D boxes with base sides at most $\delta$ and
height at most 1. Consider the MNFDH packing applied to $I$. If
MNFDH cannot place more boxes from $I$ into a bin of size $(a, b,
c)$, then either all boxes of $I$ has been packed into the bin or
the total packed volume in the bin is at least
 $
    (a-\delta)(b-\delta)(c-1).
 $
\end{lemma}
Given any feasible 3D strip packing of height $h$, we can scan a
plane parallel to the square base of the strip from the bottom to
the top to obtain a vector $x=(x_1,\ldots,x_q)$ such that
$\sum_{i=1}^{q} x_i = h$, where $q$ is the number of patterns to
pack all squares induced from the input list into a unit square bin
and $x_i$ is the height of pattern $i$.

\para{\bf Definition of  $S(K, \delta)$.} If an input set $I$ has a
constant number of different sizes, say $K$, and all the base sides
are at least $\delta$, where $\delta$ is a constant, then we define
this problem as {\em Restricted 3D strip packing with square bases},
denoted by $S(K, \delta)$.

\begin{lemma} \label{lemma:patterns}\cite{BS06}
The number of all feasible patterns of packing the square items,
induced from an instance of $S(K, \delta)$, into a unit square bin
is a constant.
\end{lemma}

\begin{lemma} \label{lemma:restrict}
$S(K, \delta)$ can be solved within $OPT+K$ in polynomial time of
$n$, where $OPT$ is the optimal cost for $S(K, \delta)$ and $n$ is
the input size.
\end{lemma}
The proof is put to the appendix.

\begin{lemma}\label{lemma:largepacking}
Assume the input set $I$ contains boxes with base sides at least
$\delta$. Then for any $K>0$, we can get a solution within
$OPT(I)(1+\frac{1}{\delta^2 K}) + K$ in polynomial time for packing
$I$ into the strip.
\end{lemma}
The proof is put to the appendix.

\paragraph{\bf Asymptotic PTAS}
Using the similar techniques as in \cite{BS06}, we present an APTAS.
Given an input set $I$ and any $\epsilon >0$, our packing is as
follows.
\begin{enumerate}
\item Let $w_j$ be the base side length of item $j$. Define
$M_i = \{j: w_j \in [\epsilon^{2^{i+1}-1},\epsilon^{2^{i}-1})\}$ for
$i = 1,...,r+1$, where $r = \lceil 1/\epsilon \rceil$.

\item Set $M:= M_i$ for some index $1 \le i \le r$ satisfying
$Vol(M_i) \le \epsilon Vol(I)$ (such a set $M_i$ must exist), where
$Vol(X)$ is the total volume of items in $X$. Define the set of
large items as  $L =\{j: w_j \ge \epsilon^{2^{i}-1}\}$ and the set
of small items as $S =\{j: w_j < \epsilon^{2^{i+1}-1}\}$.

\item Set $K= \lceil1/(\epsilon \delta^2) \rceil$ and
round all items in $L$ up into $K$ distinct sizes, $\delta
=\epsilon^{2^{i}-1} $. Then call the algorithm in Lemma
\ref{lemma:largepacking} to get an almost optimal solution.

\item Partition the unused space in the current strip into
cuboid regions and use MNFDH to pack as many squares in $S$ as
possible into the free space. Let $S^{'} \subset S$ denote the
subset of the remaining small items that could not be packed
($S^{'}$ could possibly be empty).

\item Use MNFDH to pack $M \cup S^{'}$ at the top of the current
packing in the strip.
\end{enumerate}

\begin{theorem}
Given an input set $I$ of 3D boxes with square bases, $A(I) \le
(1+12\epsilon)OPT(I) +O(K)$, where $A(I)$ is the height used by our
algorithm and $K=\epsilon^{-O(2^{\epsilon^{-1}})}$.
\end{theorem}
\begin{proof} (Sketch.) Our argument is similar as \cite{BS06}.
After Step 4, there are two cases.

\vskip 2mm\noindent  Case 1. $S^{'}$ is not empty. Then by the proof
in Section 3.4 of \cite{BS06},
   \[
   A(I) \le  Vol(I)/(1-6\epsilon)  + O(K)
        \le (1+12\epsilon)OPT(I) + O(K).
   \]
The last step follows by assuming without loss of generality that
$\epsilon \le 1/12$.

\vskip 2mm\noindent Case 2. $S^{'}$ is empty. Set $K=
1/(\epsilon\delta^2) =\epsilon^{-O(2^{\epsilon^{-1}})}$, where
$\delta =\epsilon^{2^{i}-1} $ in Step 2. By Lemma
\ref{lemma:largepacking},

\begin{equation}\label{eqn:large}
    A(L \cup S) = A(L) \le (1+\frac{1}{\delta^2 K} )OPT(I) + K
    \le (1+\epsilon)OPT(I)+O(\epsilon^{-O(2^{\epsilon^{-1}})}).
  \end{equation}
Next, we consider the cost of packing $M$ by MNFDH. Since the base
size of each item in $M$ is at most $\epsilon$, by Lemma
\ref{lemma:MNFDH},
  \begin{equation}
    \label{eqn:middle}
    A(M) \le Vol(M)/(1-2\epsilon) +1 \le \epsilon OPT(I)/(1-2\epsilon) +1.
  \end{equation}
Combining (\ref{eqn:large}) and (\ref{eqn:middle}), $A(I) \le
(1+3\epsilon)OPT(I)+O(\epsilon^{-O(2^{\epsilon^{-1}})})$.

Finally we want to note that each step in our algorithm takes
polynomial time of $n$ since $\epsilon$ is a constant.
\end{proof}

\section{Conclusions}
In this paper, we present new asymptotic upper bounds for the 3D
strip packing problems. Our results give a possible way to apply the
approaches for 1- and 2-dimensional bin packing to 3-dimensional
strip packing. It might be interesting to see if the idea can be
used to tackle higher dimensional strip packing in the general case.
Regrading the special case that items have square bases, with the
technique in the previous work on 2D bin packing and 2D strip
packing an APTAS is easily achieved. Such an approach can also be
extended to multidimensional strip packing.

 \normalsize
\newpage
\begin{center}
{\bf Appendix}
\end{center}


 \para{\bf Proof of Lemma \ref{lemma:restrict}}
 \begin{proof}
Our idea is similar with the one in \cite{KR00}. First we consider
the following LP, where $q$ is  the number of all feasible patterns
of packing the squares induced from an instance of $S(K, \delta)$
into a unit square bin, $\alpha_{ij}$ is the number of type $j$
items in pattern $i$ and $\beta_j$ is the sum of heights of type $j$
items for $1 \le j \le K$, $x_i$ is the height of pattern $i$.
 \begin{displaymath}
     \begin{array}{lll}
       \textrm{Min.  }    &\sum_{i =1}^{q} x_i &   \\
       \textrm{s.t.  }    &\sum_{i=1}^{q} \alpha_{ij} x_i \ge \beta_j,
                      \textrm{  }  &  1 \le j \le K  \\
                           & x_i \ge 0,  & 1\le i \le q .
  \end{array}
  \end{displaymath}
By Lemma \ref{lemma:patterns}, $q$ is a constant related to $K$ and
$\delta$. So, the above LP can be solved in polynomial time of $n$,
where $n$ is the input instance size of  $S(K, \delta)$. Let $x^*
=(x_1^*,...,x_2^*)$ be an optimal solution of the above LP. By some
linear programming property, there are at most $K$ non-zero
$x_i^*$'s. Up to renaming, we assume the non-zero coordinates are
$x_1^*,...,x_K^*$. We construct a packing of $S(K, \delta)$ in the
following way.

We fill in the strip bottom-up, taking each pattern in turn. Let
$x_j^* >0$ be the current pattern. Pattern $j$ will be used between
level $l_j = (x_1^* +1)+\cdots+ (x_{j-1}^*+1)$ and level
$l_{j+1}=l_j+x_j^*+1$ (initially $l_1 = 0$). For each $i$ such that
$\alpha_{ij}\ne 0$, we draw $\alpha_{ij}$ cuboids of base size $w_i$
going from level $l_j$ to level $l_{j+1}$, where $w_i$ is the base
side length of type $i$ item. After this is done for all $j$'s, we
take all the cuboids of width $w_i$ one by one in some arbitrary
order, and fill them in with the boxes of base size $w_i$ in a
greedy manner (some small amount of space may be wasted on top of
each column).

Since every box has its height at most $1$, all boxes can be packed
in the above way. Moreover $\sum x_i^*$ is a lower bound of the
optimal value for $S(K, \delta)$. Hence we have this lemma.
\end{proof}

\para{\bf Proof of Lemma \ref{lemma:largepacking}}
 \begin{proof}
The algorithm has 3 steps:
\begin{description}

\item [Stacking.]
Sort the $n$ boxes  in non-increasing order of base sizes and stack
up them one by one to get a stack of height $H$. And define $K-1$
threshold boxes, where a box is a threshold if its interior or
bottom base intersects some plane $z=\frac{iH}{K}$, for $1 \le i \le
K-1$.

\item[Grouping and rounding.]
The threshold boxes divide the remaining boxes into $K$ groups. The
base sizes of the boxes in the first group are rounded up to 1, and
the base sizes of the boxes in each subsequent group are rounded to
the base size of the threshold box below their group. This defines
an instance $I_{sup}$ of $S(K, \delta)$.

\item[Packing.] Apply the approach of Lemma \ref{lemma:restrict}
to $I_{sup}$ and output the packing.
\end{description}

To analyze the algorithm, we use the argument of Kenyon and R\'emila
\cite{KR00}. Consider two instances $I_{inf}'$ and $I_{sup}'$
derived from the stack built in stacking step. The two instances are
obtained by first cutting the threshold boxes using the planes
$z=\frac{iH}{K}$, then considering the $K$ subsequent groups of
boxes in turn (where each group now has cumulative height exactly
$H/K$); to define $I_{sup}'$, we round the base sizes in each group
up to the largest base size of the group (up to 1 for the first
group); to define $I_{inf}'$, we round the base sizes in each group
down to the largest base size of the next group (down to 0 for the
last group). Let $lin(I)$ be the solution of the above linear
programming for instance $I$. It is easy to see that
\[
  lin(I_{inf}') \le lin(I) \le lin(I_{sup}) \le  lin(I_{sup}').
\]
 Moreover,
 \[
     lin(I_{sup}') \le lin(I_{inf}')+ H/K.
\]
Since $OPT(I) \ge lin(I)$ and $OPT(I) \ge \delta^2 H$, the height
used by our packing is at most $lin(I_{sup}) + K \le  lin(I_{inf}')+
H/K + K \le OPT(I) + K + H/K \le OPT(I)(1+\frac{1}{\delta^2 K}) + K.
$
 \end{proof}
\end{document}